\documentclass{article}
\usepackage{arxiv}

\usepackage[T1]{fontenc}    
\usepackage{hyperref}       
\usepackage{url}            
\usepackage{microtype}      
\usepackage{graphicx}

\title{Photoreflectance spectroscopy of $\textrm{BiOCl}$ epitaxial thin films}

\author{ T. Nishiwaki \\
	Far-IR R\&D Center\\
	University of Fukui\\
	Fukui 910-8507, Japan\\
	\And
	\href{https://orcid.org/0000-0002-5706-8909}{\includegraphics[scale=0.06]{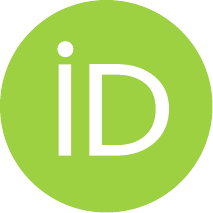}\hspace{1mm}T. Makino} \\
	Far-IR R\&D Center\\
	University of Fukui\\
	Fukui 910-8507, Japan \\
	\texttt{tmakino@u-fukui.ac.jp} \\
	\AND
	\href{https://orcid.org/0000-0002-5219-0307}{\includegraphics[scale=0.06]{orcid.pdf}\hspace{1mm}Z. Sun} \\
        SKL-ATMSP \\
	Wuhan University of Technology\\
	Wuhan 430070, China\\
	\And
	\href{https://orcid.org/0000-0003-2747-9675}{\includegraphics[scale=0.06]{orcid.pdf}\hspace{1mm}D. Oka} \\
        Department of Chemistry, \\
	Tokyo Metropolitan University\\
	Hachioji 192-0397, Japan \\
	\And
	\href{https://orcid.org/0000-0002-8957-3520}{\includegraphics[scale=0.06]{orcid.pdf}\hspace{1mm}T. Fukumura} \\
	Department of Chemistry and WPI-AIMR \\
        Tohoku University\\
	Sendai 980-8578, Japan \\
}

\renewcommand{\shorttitle}{Photoreflectance of BiOCl}

\hypersetup{
pdftitle={Photoreflectance spectroscopy of BiOCl epitaxial thin films},
pdfsubject={cond-mat},
pdfauthor={T. Nishiwaki, T.~Makino},
pdfkeywords={accepted for publication in JJAP},
}

\begin{document}
\maketitle

\begin{abstract}
We have observed a new optical transition in the photoreflectance
spectra of indirect-gap BiOCl thin films, which were grown on
SrTiO\textsubscript{3} substrates. The position of this transition is
close in energy to its bulk critical point energy. Moreover, these are
significantly lower than a higher-lying direct-type critical point from
an energetic point of view. The spectral line shape analysis for our
observed signal suggests the presence of an excitonic effect of this
compound. We determined its dependence of the optical anomaly on
temperature ranging from 80 K to room temperature. We adopted the
Varshni model for this analysis. At last, we compared photonic
properties of BiOCl with those of an element and binary semiconductors.
\end{abstract}

\section{Introduction}
Bismuth oxychloride (BiOCl) is one of the oxyhalide compounds. This
compound belongs to the P4/nmm, \emph{i.e.}, a tetragonal structure.
Namely, its crystal structure can be regarded as the same as that of
matlockite ~{[}1--4{]}. This has played an important role in both basic
and applied perspectives as one of the wide-gap materials because of its
potential photocatalytic and optoelectronic applications ~{[}5,6{]}.

Despite the abovementioned potentials, even a modern deposition
technique has not been able to accomplish a flat surface ~{[}7{]}, owing
to its low symmetry. Until now, a number of attempts at thin film growth
have shown that polycrystalline nanoplatelets or nanoflower-like
particles have been formed ~{[}8{]}. This is also true for the
conventional synthesis of bulk crystals ~{[}9{]}. To the best of our
knowledge, no previous study has investigated the detailed optical
properties, probably because the forms of the previously produced
specimens are not adequate for the studies on the intrinsic physical
properties. We here point out a pivotal work to attack this problem. Sun
\emph{et al}. have recently grown successfully epitaxial layers of BiOCl
with a flat surface and high crystallinity. In their work, they adopted
a mist chemical vapor deposition (CVD) method~{[}7{]}. In that study,
careful controls on various growth conditions such as carrier gas flow
rate, growing temperature, and choice of an appropriate substrate
(SrTiO$_3$) allowed them to grow specimens with a flat
surface, which potentially facilitates the detailed characterization of
their intrinsic properties.

Several researchers so far have conducted first-principles calculations
on the electronic structures of this oxyhalide. As earlier-mentioned, it
is shown that the oxyhalides are classified as indirect-type
semiconductors with the energy gap of \emph{ca.} 2.9 eV~{[}10{]}. We
have considered that the optical response at the bandedge region of
indirect semiconductors is weak. On this subject, much less is known
about its properties. Especially in the case of thin films, the volume
of the specimen under test cannot be useful to cover the weakness of
such a bandedge-response-related signal. Therefore, observation of such
a response in the case of indirect-type semiconductors is rarely
expected when the specimen is only available in the form of thin films
or epilayers. As earlier-mentioned, the grade suitable for detailed
optical characterization having an atomically flat surface is only
available in the form of thin films, which has discouraged the
researchers from studying the spectroscopic study of the indirect-type
semiconductors at the bandedge region. Recently, however, large optical
transitions at bandedge have been observed in hexagonal-BN and diamond,
both of which are albeit indirect semiconductors~{[}11--13{]}. In these
materials, both the strong excitonic effects and the strong
electron-phonon interactions give rise to the observation of
sufficiently strong optical transitions.

Despite not being intrinsic, another recent observation is the formation
of an interface-related midgap state situated near the indirect-bandedge
of TiO$_2$ grown on F-doped SnO$_2$, which
leads to an optical response near the energy of the
indirect-bandedge~{[}14{]}.

Even more, a Chinese group recently reported the results of theoretical
calculations of an interesting electronic structure of
oxyhalides~{[}15{]}. They reported that the formation of
heterostructures between oxyhalides would turn an indirect-type into a
direct-type semiconductor. Therefore, it could be expected that BiOCl
grown on SrTiO$_3$ also becomes a direct-type semiconductor
due to its heterostructure nature, and hence a strong optical response
is observed at the indirect-bandedge.

As earlier mentioned, there are several possibilities leading to us the
observation of the strong optical transitions even at the
indirect-bandedge region. However, few examples of research have been
conducted from such a point of view in the oxyhalides, which might be
promising to gain understanding or deeper insight into their physical
properties.

We paid attention to one of the spectroscopic tools in this work. This
is known as photoreflectance (PR) spectroscopy. This measures the
differential reflectance under the condition of applying an
additional-irradiated-laser-beam- modulated electric field to the device
under test (DUT)~{[}16{]}. Because of these features, nondestructive and
non-catalytic measurements can be attained by PR spectroscopy. In
addition, the PR is known to have an energy-derivative and
background-free nature. Therefore, this is considered appropriate for
the detection of the bandedge optical transition.

To strengthen the potentials of BiOCl, it is necessary to further
elucidate the fundamental properties from a material scientific point of
view. For example, we could come up with a temperature-dependent optical
property as one of the fundamental and technically essential properties
of semiconductors. Expectedly, one can obtain through this study further
insights into various aspects such as electron-phonon interaction,
thermophysical phenomenon, and photonic properties. A search of the
literature revealed that spectroscopic measurements have been performed
only at room temperature up until now. There is almost no published data
on the cryogenic temperature region.

In the current paper, we report on the discovery of a new optical
transition in BiOCl thin films grown on SrTiO$_3$
substrates, assessed with PR spectroscopy. We also estimated the
temperature dependence of their optical properties. Special emphasis is
placed on the interpretation of the temperature dependence of the
bandgap energies with respect to the electron-phonon interaction.
Finally, we compare the fit-determined parameters with those of other
elements and binary semiconductors.

\section{Experimental methods}
\subsection{Sample preparation}

We grew epitaxial BiOCl thin films at a temperature of 350 $^{\circ}$C. We
adopted SrTiO$_3$ (STO) as a substrate. There is a
considerable difference in the lattice constants between BiOCl and STO,
which was estimated to be \emph{ca.} 0.33\%. We grew the thin film, the
typical thickness of which was \emph{ca.} 260 nm. In this study, we
adopted the mist-CVD technique~{[}17,18{]}. It is now well known that
this facile-solution-based technique enables fundamental growth even
under atmospheric pressure. During the growth, we utilized
N,N-dimethylformamide (DMF, purity: 99.5\%; Wako) solutions of
BiCl$_3$ (purity: 99.99\%; Kojundo Chemical Lab., 0.02
mol/L) as a precursor solution. This is followed by nebulization of the
precursor solution into microscale mist and transfer to the substrates
in a tubular furnace with N$_2$ gas from the carrier and
dilution ports at 3.0 L/min and 0.5 L/min, respectively ~{[}7{]}.

\hypertarget{photoreflectance-spectroscopy}{%
\subsection{Photoreflectance
spectroscopy}\label{photoreflectance-spectroscopy}}

Here, we explain an optical characterization method. The present study
utilizes PR spectroscopy to analyze our observed data. Namely, we used a
325 nm line from a HeCd laser as a pumping source and monochromatized
light from a Xe arc lamp as probe. Then, we attached DUT's on a cold
finger of a refrigerator for measurements in the cryogenic temperature
range.

In this study, we used the normalization technique with careful
treatment to avoid noise and spurious contributions. We have given a
detailed explanation of why we adopted this technique
elsewhere~{[}14,19--22{]}. We summarize briefly here: we chose the
DC-normalization procedure among several techniques, such as
beam-sweeping and dual-chopping schemes. For the elimination of unwanted
long-term thermal drift, great care was taken in the adoption of digital
electronic components~{[}22{]}. Throughout this study, we show the
results after the facile \emph{a-posteriori} subtraction of the constant
background~{[}23,24{]}.

\hypertarget{calculation-procedures}{%
\section{Calculation procedures}\label{calculation-procedures}}

To calculate the band structures of BiOCl, we used the plane wave basis
set PWscf package of Quantum ESPRESSO (QE)~{[}25{]},
an~\emph{ab-initio}~density functional theory program with the plane
wave basis, and a pseudopotential method~{[}26--29{]}. In addition to
calculating band dispersion, QE can also perform structural optimization
calculations, eigenfrequencies, and dielectric constants.
First-principles calculations within the framework of density functional
theory (DFT)~{[}30{]} is performed to analyze the structural and
electronic properties of BiOCl. In this calculation, we neglected the
contribution from SrTiO$_3$. For the atomic coordinates of
BiOCl, the structure that has been adopted in the previous DFT
calculations, consistent with the experimental results, was adopted. To
find the influence of the electron density on the exchange correlation
energies of ions, we have used the LDA functional, which belongs to the
class of Methfessel-Paxton functionals. The Brillouin zone (BZ) was
integrated using an 8 $\times$ 8 $\times$ 8 centered Monkhorst-Pack \emph{k}-point
grid. As a precaution, it is necessary to adopt a value for which the
cutoff energy converges. Our cutoff energy is \(1 \times 10^{-8}\) Ry
(\(1.36 \times 10^{-7}\) eV). The Bi, Cl, and O atoms are represented
by norm-converving pseudopotentials, and the kinetic energy and charge
density cutoffs are chosen to be 50 and 400 Ry (680 and 5442 eV),
respectively. Methfessel-Paxton smearing of the Fermi-Dirac
distribution, with a smearing width of 0.02~Ry (0.272 eV).

\hypertarget{theoretical-models}{%
\section{Theoretical models}\label{theoretical-models}}

\hypertarget{electronic-energy-band-structures-of-biocl}{%
\subsection{Electronic energy-band structures of
BiOCl}\label{electronic-energy-band-structures-of-biocl}}

For purposes of discussion, detailed information on the electronic
energy-band structures can be obtained with a wide variety of
calculations of BiOCl. We also calculated the electronic energy-band
structure of this compound. Figure 1 shows the calculated band
structures of BiOCl along the high symmetry directions of the Brillouin
zone (BZ). As can be seen, the crystal has an indirect bandgap. In other
words, the fundamental absorption edge of this compound corresponds to
the indirect bandgap. As indicated with an arrow in the figure, the
optical transition is labeled \emph{E}$_0$. The top of the
valence band (VB) is at the A point, and the bottom of the conduction
band (CB) is at the Z point, respectively.

\begin{figure}
	\centering
	\includegraphics[width=0.85\linewidth]{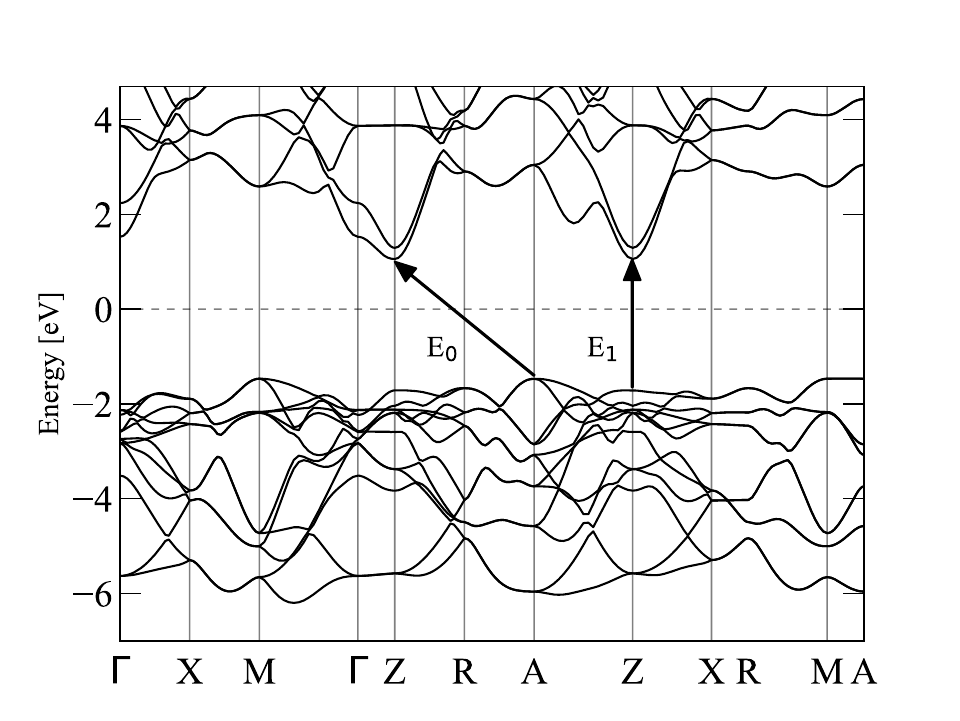}
	\caption{Electronic energy-band structure of BiOCl.}
\end{figure}

The calculated indirect DFT bandgap for BiOCl is \emph{ca.} 2.5 eV. We
can also confirm the higher-lying direct gap of \emph{ca.} 2.8 eV. As
shown with an arrow in the figure, the corresponding optical transition
is labeled \emph{E}$_1$. It is well known that in DFT
calculations, the bandgap of an insulator is often
underestimated~{[}31{]}. Our calculated value of the indirect bandgap
energy is even lower than the ones calculated by other groups~{[}10{]}
as large as 2.9~eV. This is probably because we did not account for the
Van-der-Waals effect in our calculation. This is not a serious concern
because we show this band diagram for the main purpose of explaining
optical transitions and assignments.

Considering the typical DFT-related `scissor' shift ranging several
hundreds of milli-electron-volts and the results of experimental studies
conducted by other groups, the energy of the indirect-transition bandgap
should be 3.2~eV in this compound. This agrees with the energetic range
of our observed signal. As later discussed in detail, it is difficult to
assign the PR signal to the indirect bandgap response. Hence, we
tentatively assign this to an exciton related to the higher-lying
\emph{E}$_1$ gap (\emph{ca.} 3.5~eV).

\hypertarget{standard-critical-point-theory}{%
\subsection{Standard critical point
theory}\label{standard-critical-point-theory}}

We here describe the model adopted throughout this study. This is called
a standard critical point (SCP) model. Cardona, Aspens and others have
established this critical point modeling~{[}32,33{]}. This model relies
on the experimentally obtained knowledge that fundamental features in
the optical bandedge response are due to interband transition around the
critical points. We analyzed our PR spectra obtained in this study are
analyzed using the SCP model~{[}34{]}. Based on this model, letting us
set \emph{C}, \(\theta\), and \(\Gamma\) as an amplitude, phase angle
and damping-related parameter, the differential reflectivity simply
reads itself:

\begin{equation}
\frac{\Delta R}{R}= Re \left(-Ce^{i\theta}\left(E - E_{1} + {i\Gamma} \right)^{-n} \right), 
\end{equation}

where \(\Delta R\) is reflectivity change, $R$ is reflectance, $E$ is
photon energy, and \(E_{1}\) is the band gap energy at each temperature.
Also, n represents the type of optical transition: \(n\) = 2, 2.5, and 3
for an excitonic transition, a three-dimensional (3D) one-electron
transition, and a two-dimensional one-electron transition, respectively.
The phase angle \(\theta\) takes the influence of an inhomogeneous
electric modulation field into account~{[}34{]}.

\section{Results and discussions}\label{results-and-discussions}

Figure 2 shows the measured PR spectra (solid lines) and the results of
the fit based on the SCP theory to the measured data (dashed lines).
These are taken for BiOCl thin films at temperatures of 100 to 300 K.
Here, $E_1$ was estimated from SCP theory. It should be noted that, to the
best of our knowledge, the PR signal associated with indirect transition
has not been reported even for another indirect semiconductor such as
SiC. Therefore, we exclude this assignment related to the
\emph{E}$_0$ critical point. Alternatively, we assign this
anomaly to the direct-type transition from the valence band of BiOCl to
the interface midgap state, as has been observed in
TiO$_2$/F-SnO$_2$ heterosystem~{[}14{]}. This
is because many years of research on harmonic generation have
established the picture that optical processes through the midgap level
almost always involve the true midgap level, and we believe that the
midgap level is also the true midgap level in our case~{[}35{]}.

\begin{figure}
	\centering
\includegraphics[width=0.6\linewidth]{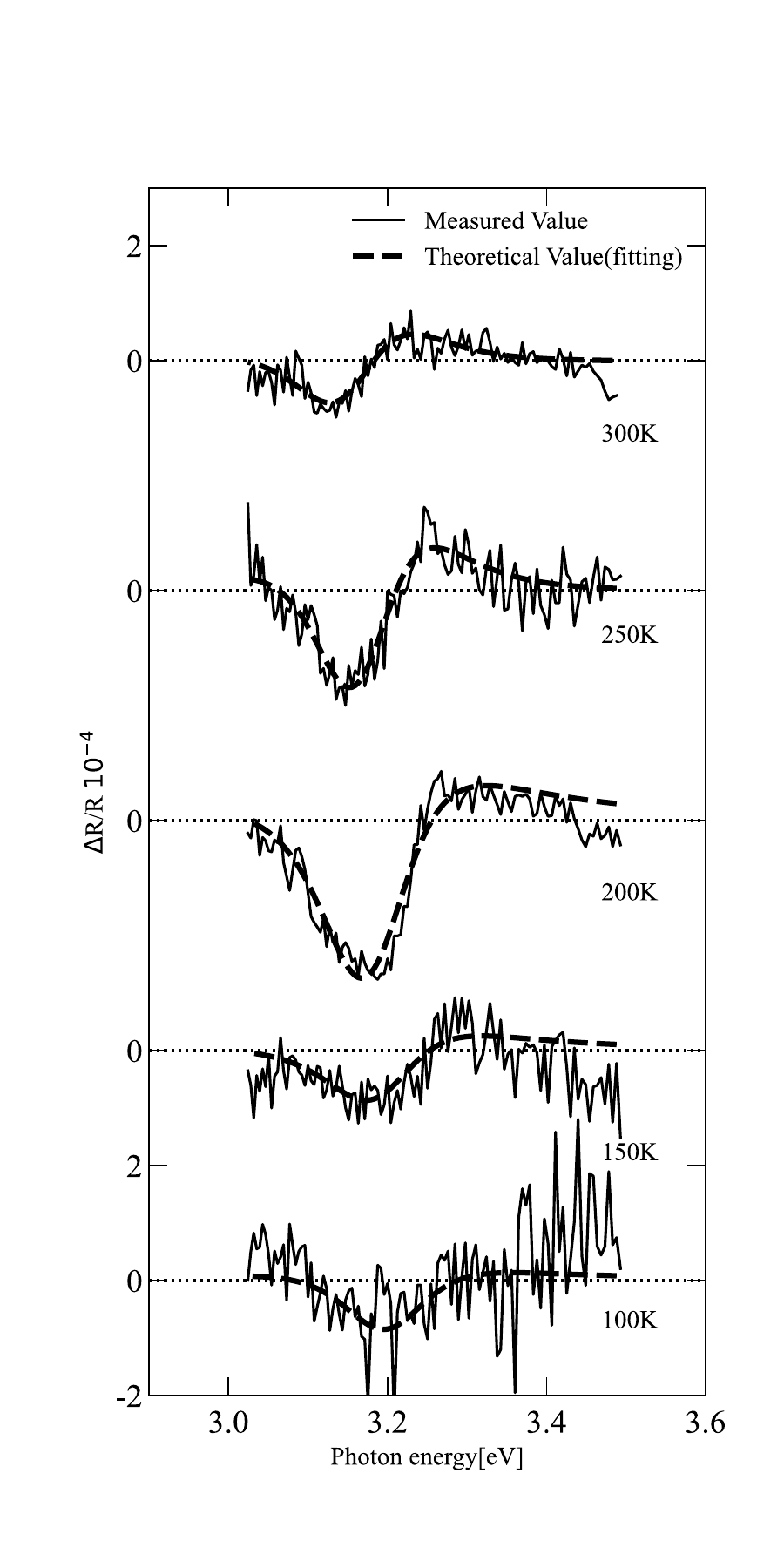}
\caption{The photoreflectance spectra of BiOCl thin film measured at different temperatures along with the calculated results based on SCP theory (dashed lines). The solid lines are the experimental data.}
\end{figure}

\begin{figure}
	\centering
\includegraphics[width=0.75\linewidth]{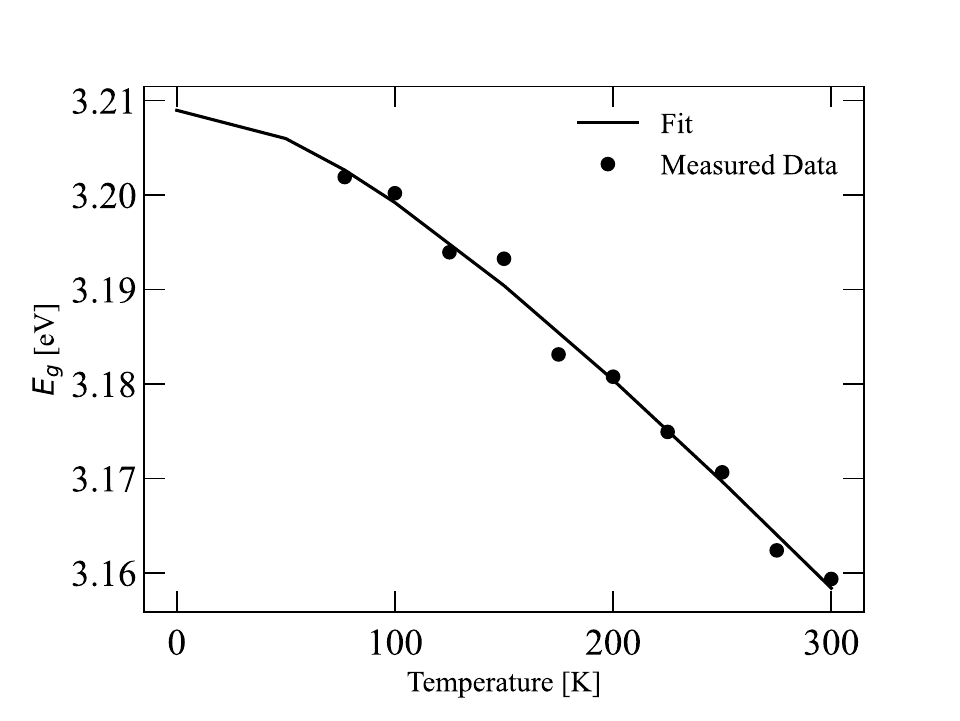}
\caption{Temperature dependences of direct bandgap $E_1$ observed in BiOCl (closed symbols). A curve (solid line) corresponds to a numerical fit to the experimental data field using eq. (2) with phonon dispersion-related parameters.}
\end{figure}

\begin{table}
	\caption{Fitting parameters based on SCP theory at temperatures of 100 K to 300 K}
	\centering
\begin{tabular}{ccccc}
\hline
Temperature (K) & $E_1$ (eV) & $\Theta$ (rad) & $C$ & $\Gamma$
(meV)\\
\hline
100 & 3.200 & 3.292 & $8.9\times10^{-7}$ & 102\\
150 & 3.193 & 3.705 & $1.1\times10^{-6}$ & 108\\
200 & 3.181 & 3.504 & $3.3\times10^{-6}$ & 108\\
250 & 3.171 & 5.318 & $3.1\times10^{-7}$ & 121\\
300 & 3.159 & 5.722 & $1.6\times10^{-7}$ & 124\\
\hline
\end{tabular}
\end{table}

As seen in Table~1, the energy positions of the \emph{E}$_1$
peaks gradually increase as the temperature decreases from 300 to 100 K.
For temperatures lower than 100 K, the energy positions of these peaks
do not change so clearly (not shown in the figure). A similar tendency
is observed for the spectral widths of these peaks. The dashed lines
represent the theoretical fits using eq. (1). The excitonic (\(n = 2\))
transition is taken into consideration in these fits. We successfully
evaluated these regression parameters to be strength parameter
$C =7\times 10^{-6}$ and broadening parameters
\(\Gamma = 106\) meV at 80 K, respectively. Other regression parameters
are also summarized in Table 2. The abovementioned excitonic assignment
favors our conjecture related to the \emph{E}$_1$ direct
bandgap. In this case, we must attribute the energetic difference to the
exciton binding energy, which amounts to \emph{ca.} 300~meV. This is
exceptionally large for the exciton binding energy of inorganic
materials, although literature data reported the binding energy
exceeding 200~meV for a delafossite-type cuprous oxide~{[}36{]}. In
addition, there is no published data suggesting presence of a sizable
excitonic effect in oxyhalides. As has been extensively discussed in
perovskite lead halides, it is now well known that the exciton binding
energy severely depends on the crystalline shape. The microcrystals tend
to exhibit smaller binding energies. So far, only the formation of
polycrystalline nanoplatelets or nanoflower-like particles has been
reported in oxyhalides. This is not advantageous due to the sizable
excitonic effect. In other words, the availability of
single-crystalline-grade epilayers can have the possibility of a sizable
excitonic effect, which is suggestive for the consistent interpretation
of our observations. On the other hand, let us discuss the assignment
related to an indirect exciton, which has been recently observed in BN
and diamonds as a strong optical transition even in the
indirect-bandedge region. It is not necessary to consider the
exceptionally large exciton binding energy for oxyhalides. It is
considered that this assignment is, however, less plausible because this
effect is cooperative, resulting both from a sizable excitonic effect
and from an exceptionally large electron-phonon interaction. Moreover,
this has only been reported for super-wide-gap semiconductors, which is
not the case for oxyhalides. Nevertheless, extensive work should be
conducted in the future for a more plausible and detailed assignment.

In Fig. 3, the plots of the critical point (CP) energies, \(E_{1}\)
versus temperature determined from the SCP fit analysis are shown.
Closed circles in Fig. 3 depict the bandgap energy (\(E_{1}\)) in BiOCl
as a function of temperature. We confirm that the bandgap energy
(\(E_{1}\)) is monotonically decreasing along with the increase in
temperature (\emph{T}). Because the repulsion between electrons plays a
role in the determination of the bandgap energy, the bandgap shrinkage
is certainly a result of the thermal expansion of the crystal volume.

For comparison purposes, we adopted Varshini's model to analyze the
temperature-dependent bandgap energy, although there is a more flexible
model~{[}37,38{]}. Here, the corresponding temperature-dependent
equation can be expressed as:

\begin{equation}
E_{1}\left( T \right) = E_{1}\left( 0 \right) - \frac{\alpha T^{2}}{T + \beta}, 
\end{equation}

where $E_{1}(0)$ is the bandgap energy at $T=0$ K,
$\alpha$ and $\beta$ are empirical and more or less phenomenological
parameters~{[}39{]}. The line in Fig.~3 denote the result of
least-squares fits of eq. (2) to the experimental data~{[}38{]}. The
parameters are evaluated to be \(\alpha = 2.6 \times 10^{-4}\)~eV/K,
\(\beta =170\)~K, \(E_{1}\left( 0 \right)=3.21\)~eV, respectively. The
calculated result (solid line in Fig.~3) is in reasonably good agreement
with the experimental tendency, which is monotonic.

\begin{table}
	\caption{Compilation of the Varshni-related parameters for
various elemental and compound semiconductors~{[}37{]}. The units of
$\alpha$ and $\beta$ are electron volt per Kelvin and Kelvin,
respectively.}
	\centering
	\begin{tabular}{cccc}
\hline
Substance & $E_1(0)$ (eV) & $\alpha$ (eV/K) & $\beta$ (K)\\ \hline
Si & 1.16 & $7.0\times 10^{-4}$ & 1100\\
Ge & 0.741 & $4.6\times 10^{-4}$ & 210\\
GaAs & 1.53 & $8.9\times 10^{-4}$ & 570\\
Diamond & 5.42 & $-2.0\times 10^{-4}$ & -1440\\
BiOCl & 3.21 & $2.6\times 10^{-4}$ & 170\\
\hline
	\end{tabular}
\end{table}

We listed the Varshni-model-related parameters of a variety of elemental
and compound semiconductors in Table~2 for comparison. We took these for
Si, Ge, GaAs, and diamond from Ref.~{[}37{]}. Although these parameters
have been regarded as more or less phenomenological parameters, one of
the hypothetical works insisted that \(\beta\) is closely related to
Debye's temperature. Experimentally, the $\beta$ values of BiOCl are
particularly low among the enumerated substances. It is considered that
it is related to the relatively low optical phonon energy inherent to
the Van-der-Waals bond nature. Because the oxyhalides are famous for the
Van-der-Waals effect as a bonding mechanism, the relatively low
concentration \(\beta\) seems to be consistent with this conjecture.

The effect of temperature has an influence not only on the energetic
shift of the critical point (CP) but also on its broadening
\(\left( \Gamma \right)\). Ozaki and coworkers have recently formulated
a model for the purpose of temperature-induced broadening by taking into
account the Bose-Einstein occupation factor~{[}40,41{]}. By assuming
that the electron (exciton)--optic phonon results in the broadening, and
letting \(\Gamma_{1}\), \(\Gamma_{0}\), and \(\Theta_{\Gamma}\) as a
temperature independent broadening, a phenomenological parameter, and an
effective phonon temperature, its equation reads itself:

\begin{equation}
\Gamma\left( T \right) = \Gamma_{1} + \frac{\Gamma_{o}}{\exp\left( \frac{\Theta_{\Gamma}}{T} \right) - 1}.
\end{equation}

As an example of such a temperature-independent broadening mechanism, we
can point out the Auger process, electron-electron interaction,
crystalline imperfections, and surface scattering~{[}40,41{]}. This
equation tells us about an approximately constant value of \(\Gamma\)
from a low temperature of 80 K up to \emph{ca.} 100 K. A
temperature-dependent term in eq. (3) can no longer be negligible, which
is followed a proportional increase at temperatures higher than
\emph{ca.} 100 K. The solid line in Fig. 4 correspond to the result of a
fit to the experimentally obtained \(\Gamma\) parameters. In the
remainder, we compare an effective phonon temperature
\(\Theta_{\Gamma}\) with the reported optic phonon energies of this
compound. The maximum value of the optic phonon energy is \emph{ca.}
25~meV, which is equivalent to 290 K ~{[}42{]}. This is significantly
low compared with the value of \(\Theta_{\Gamma}\), which is 360 K. It
is believed so far that only one kind of optic phonon or weighted
contribution from several phonons determines this effective
temperature. On the contrary, our experimental result suggests rather
that concomitant contributions both from \emph{E}$_g$
(7.3~meV) and \emph{A}$_{1g}$ (25~meV) phonons, because the
sum of these phonon energies coincide with the value of
\(\Theta_{\Gamma}\) (360 K or equivalently 31~meV). Although more
extensive works are necessary to support our preliminary assignment in the future, the energetic coincidence with the experimental result seems to
capture the main features observed here. As has been demonstrated in
lasing action mechanism in a certain class of semiconductor
nanostructures, the importance of the polaron effect which is a
consequence of the electron-phonon interaction is evident. Thus, the
polaron-related material parameters obtained in this work are expected
to be useful for future optoelectronic applications.

\begin{figure}
	\centering
\includegraphics[width=0.7\linewidth]{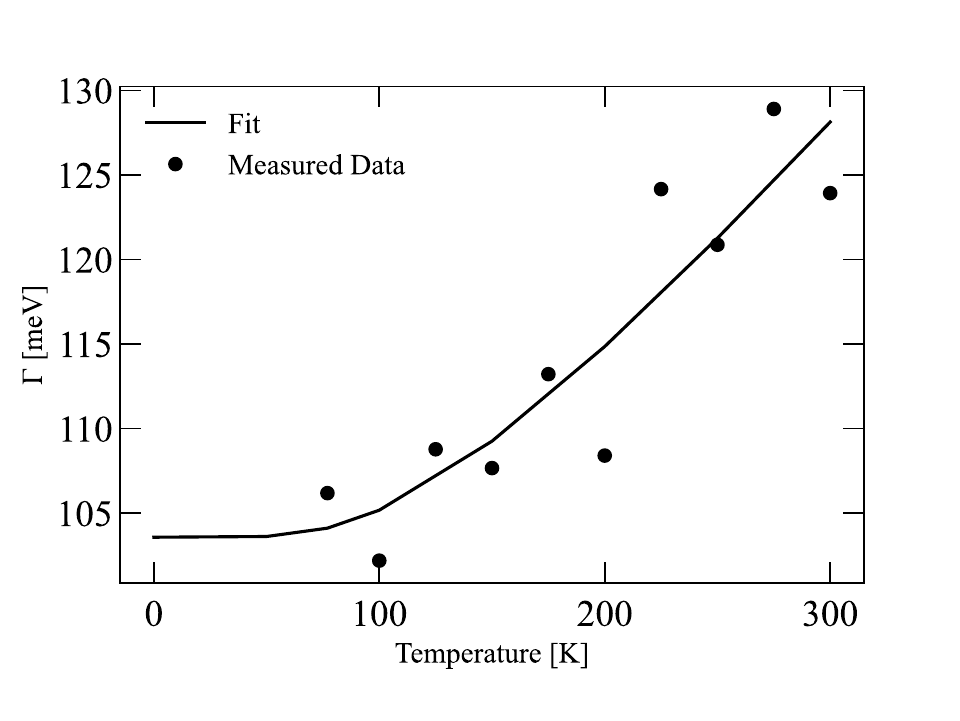}
\caption{Temperature dependence of the broadening parameter $\Gamma$ (closed circles). The solid line represents the best-fit results of $\Gamma$(T) with eq. (3). The fit-determined parameters are $\Gamma_1= 103$ meV, $\Gamma_0=57$ meV, and $\Theta_\Gamma = 360$ K, respectively.}
\end{figure}

\section{Conclusion}

We assessed the temperature dependence of the optical properties of
BiOCl thin films. We have been able to extract several important
material parameters. We interpreted the TDBGE using a model based on the
electron-phonon interaction. The temperature dependence of the CP
parameters has been determined and analyzed using the Varshni
equation~{[}37{]} and an alternative analytical formula developed by
Ozaki \emph{et al.} recently~{[}41{]}. We also compared the parameters
with other substances using the Varshni equation. In particular, low
values of $\beta$ were found, which may be related to the relatively low
optical phonon energy inherent in the nature of van der Waals coupling.
Also, he polaron-related material parameters obtained in this work are
expected to be useful for future optoelectronic applications.

\hypertarget{acknowledgments}{%
\section*{Acknowledgments}\label{acknowledgments}}

A partial financial support of this work from JSPS KAKENHI Grant Number
19K05303 was also acknowledged. We acknowledge technical assistance from
E. Kobayashi.

\section*{References}
\begin{enumerate}
\item K. T. Drisya, S. Cort\'es-Lagunes, A.-L. Gardu\~{n}o-Jim\'enez, R. N. Mohan, N. Pineda-Aguilar, A. C. Mera, R. Zanella, and J. C. Dur\'an-Alvarez, J. Environ. Chem. Eng. \textbf{10}, 108495 (2022).
\item K. G. Keramidas, G. P. Voutsas, and P. I. Rentzeperis, Zeitschrift f\"ur Krist. \textbf{205}, 35 (1993).
\item W. L. Huang and Q. Zhu, Comput. Mater. Sci. \textbf{43}, 1101 (2008).
\item Q. Wang, J. Hui, Y. Huang, Y. Ding, Y. Cai, S. Yin, Z. Li, and B. Su, Mater. Sci. Semicond. Process. \textbf{17}, 87 (2014).
\item Y. Yang, C. Zhang, C. Lai, G. Zeng, D. Huang, M. Cheng, J. Wang, F. Chen, C. Zhou, and W. Xiong, Adv. Colloid Interface Sci. \textbf{254}, 76 (2018).
\item L. Zhang, Z. Han, W. Wang, X. Li, Y. Su, D. Jiang, X. Lei, and S. Sun, Chem. Eur. J. \textbf{21}, 18089 (2015).
\item Z. Sun, D. Oka, and T. Fukumura, Chem. Commun. \textbf{56}, 9481 (2020).
\item D. S. Bhachu \textit{et al.}, Chem. Sci. \textbf{7}, 4832 (2016).
\item L. S. G\'omez-Vel\'azquez, A. Hern\'andez-Gordillo, M. J. Robinson, V. J. Leppert, S. E. Rodil, and M. Bizarro, Dalton Trans. \textbf{47}, 12459 (2018).
\item L. Zhang, Z.-K. Tang, W.-M. Lau, W.-J. Yin, S.-X. Hu, and L.-M. Liu, Phys. Chem. Chem. Phys. \textbf{19}, 20968 (2017).
\item B. Arnaud, S. Lebegue, P. Rabiller, and M. Alouani, Phys. Rev. Lett. \textbf{96}, 26402 (2006).
\item G. Cassabois, P. Valvin, and B. Gil, Nat. Photonics \textbf{10}, 262 (2016).
\item K. Konishi and N. Naka, Phys. Rev. B \textbf{104}, 125204 (2021).
\item E. Kobayashi, S. Shinmura, S. Ito, and T. Makino, Jpn. J. Appl. Phys. \textbf{59}, SCCB23 (2019).
\item J. Wang, Y. Huang, J. Guo, J. Zhang, X. Wei, and F. Ma, J. Solid State Chem. \textbf{284}, 121181 (2020).
\item J. Misiewicz, P. Sitarek, G. Sek, and R. Kudrawiec, Mater. Sci. Pol. \textbf{21}, 263 (2003).
\item G. Yong Chung, H. Dong Kim, B. Tae Ahn, and H. Bin Im, Thin Solid Films \textbf{232}, 28 (1993).
\item T. Kawaharamura, G. T. Dang, and M. Furuta, Jpn. J. Appl. Phys. \textbf{51}, 2 (2012).
\item Y. Borensztein, R. Alameh, and M. Roy, Phys. Rev. B \textbf{48}, 14737 (1993).
\item T. Kita, M. Yamada, and O. Wada, Rev. Sci. Instrum. \textbf{79}, 46110 (2008).
\item U. Behn, A. Thamm, O. Brandt, and H. T. Grahn, J. Appl. Phys. \textbf{90}, 5081 (2001).
\item G. K. Boschloo, A. Goossens, and J. Schoonman, J. Electrochem. Soc. \textbf{144}, 1311 (1997).
\item H. Shen, P. Parayanthal, Y. F. Liu, and F. H. Pollak, Rev. Sci. Instrum. \textbf{58}, 1429 (1987).
\item D. Yan, H. Qiang, and F. H. Pollak, Rev. Sci. Instrum. \textbf{65}, 1988 (1994).
\item P. Giannozzi \emph{et al.}, J. Phys. Condens. Matter \textbf{21}, 395502 (2009).
\item X. Shan and H. Chen, Phys. Rev. E \textbf{47}, 1815 (1993).
\item X. Shan and H. Chen, Phys. Rev. E \textbf{49}, 2941 (1994).
\item X. Shan and G. Doolen, J. Stat. Phys. \textbf{81}, 379 (1995).
\item X. Shan and G. Doolen, Phys. Rev. E \textbf{54}, 3614 (1996).
\item P. Hohenberg and W. Kohn, Phys. Rev. \textbf{136}, B864 (1964).
\item J. Sun, C.-W. Lee, A. Kononov, A. Schleif\'e, and C. A. Ullrich, Phys. Rev. Lett. \textbf{127}, 77401 (2021).
\item P. Lautenschlager, M. Garriga, S. Logothetidis, and M. Cardona, Phys. Rev. B \textbf{35}, 9174 (1987).
\item D. E. Aspnes and A. A. Studna, Phys. Rev. B \textbf{27}, 985 (1983).
\item D. E. Aspnes, Surf. Sci. \textbf{37}, 418 (1973).
\item W. E. Angerer, N. Yang, A. G. Yodh, M. A. Khan, and C. J. Sun, Phys. Rev. B \textbf{59}, 2932 (1999).
\item H. Hiraga, T. Makino, T. Fukumura, H. Weng, and M. Kawasaki, Phys. Rev. B \textbf{84}, 41411 (2011).
\item Y. P. Varshni, Physica (Utrecht) \textbf{34}, 149 (1967).
\item R. P\"assler, E. Griebl, H. Riepl, G. Lautner, S. Bauer, H. Preis, W. Gebhardt, B. Buda, D. J. As, D. Schikora, K. Lischka, K. Papagelis, and S. Ves, J. Appl. Phys. \textbf{86}, 4403 (1999).
\item I. A. Vainshtein, A. F. Zatsepin, and V. S. Kortov, Phys. Solid State \textbf{41}, 905 (1999).
\item L. Vi\~{n}a, S. Logothetidis, M. Cardona, Phys. Rev. B \textbf{30}, 1979 (1984).
\item S. Ozaki, T. Mishima, and S. Adachi, Jpn. J. Appl. Phys. \textbf{42}, 5465 (2003).
\item A. Rulmont, Spectrochim. Acta A \textbf{30}, 311 (1974).
\end{enumerate}
\end{document}